\documentclass[twocolumn]{aastex62}

\usepackage{amsmath}
\usepackage[T1]{fontenc}  
\usepackage{graphicx}
\usepackage{float}

\providecommand{\keywords}[1]
{
  \small	
  \textbf{\textit{Keywords---}} #1
}

\begin{document}

\title{FIRST\,J153350.8$+$272729: the radio afterglow of a decades-old tidal disruption event}


\author[0000-0002-7252-5485]{Vikram Ravi}
\affiliation{Cahill Center for Astronomy and Astrophysics, California Institute of Technology, Pasadena, CA 91125, USA}

\author{Hannah Dykaar}
\affiliation{David A. Dunlap Department of Astronomy and Astrophysics, University of Toronto \\ 50 St. George Street, Toronto, Ontario, M5S 3H4 Canada}
\affiliation{Dunlap Institute for Astronomy and Astrophysics, University of Toronto, 50 St. George St., Toronto, ON M5S 3H4, Canada}

\author{Jackson Codd}
\affiliation{Cambridge Rindge and Latin School, Cambridge, MA, USA}
\affiliation{Department of Physics, Macalester College, Saint Paul, MN 55105, USA}

\author{Ginevra Zaccagnini}
\affiliation{Cambridge Rindge and Latin School, Cambridge, MA, USA}

\author{Dillon Dong}
\affiliation{Cahill Center for Astronomy and Astrophysics, California Institute of Technology, Pasadena, CA 91125, USA}

\author{Maria R. Drout}
\affiliation{David A. Dunlap Department of Astronomy and Astrophysics, University of Toronto \\ 50 St. George Street, Toronto, Ontario, M5S 3H4 Canada}
\affiliation{Observatories of the Carnegie Institution for Science, 813 Santa Barbara St., Pasadena, CA 91101, USA}

\author[0000-0002-3382-9558]{B.~M.~Gaensler}
\affiliation{Dunlap Institute for Astronomy and Astrophysics, University of Toronto, 50 St. George St., Toronto, ON M5S 3H4, Canada}
\affiliation{David A. Dunlap Department of Astronomy and Astrophysics, University of Toronto \\ 50 St. George Street, Toronto, Ontario, M5S 3H4 Canada}

\author{Gregg Hallinan}
\affiliation{Cahill Center for Astronomy and Astrophysics, California Institute of Technology, Pasadena, CA 91125, USA}

\author{Casey Law}
\affiliation{Cahill Center for Astronomy and Astrophysics, California Institute of Technology, Pasadena, CA 91125, USA}

\begin{abstract}
  We present the discovery of the fading radio transient FIRST\,J153350.8+272729. The source had a maximum observed 5-GHz radio luminosity of $8\times10^{39}$\,erg\,s$^{-1}$ in 1986, but by 2019 had faded by a factor of nearly 400. It is located 0.15\arcsec~from the center of a galaxy (SDSS\,J153350.89$+$272729) at 147\,Mpc, which shows weak Type II Seyfert activity. We show that a tidal disruption event (TDE) is the preferred scenario for FIRST\,J153350.8+272729, although it could plausibly be interpreted as the afterglow of a long-duration $\gamma$-ray burst. This is only the second TDE candidate to be first discovered at radio wavelengths. Its luminosity fills a gap between the radio afterglows of sub-relativistic TDEs in the local universe,  and relativistic TDEs at high redshifts. The unusual properties of FIRST\,J153350.8+272729 (ongoing nuclear activity in the host galaxy, high radio luminosity) motivate more extensive TDE searches in untargeted radio surveys.   
\end{abstract}

\keywords{AGN host galaxies --- black hole physics --- radio transient sources --- time domain astronomy}

\section{Introduction} \label{sec:intro}


The ongoing Karl G.  Jansky  Very  Large  Array (VLA) Sky Survey \citep[VLASS;][]{VLASS} is the first radio all-sky survey that is sensitive to several classes of slowly evolving extragalactic radio transients, including flares from active galactic nuclei (AGN), core-collapse supernova afterglows, the orphan afterglows of off-axis $\gamma$-ray bursts, and tidal disruption events (TDEs) of stars by supermassive black holes (SMBHs). VLASS is an interferometric survey in the 2--4\,GHz band of the entire 33,885\,deg$^{2}$ of sky north of a declination of $-40^{\circ}$, with an angular resolution of 2.5\arcsec. When complete, the sky will be surveyed over three epochs spaced by 32 months, to a continuum-image rms of $120\,\mu$Jy per beam per epoch. The exceptional survey grasp of VLASS provides the opportunity to assemble samples of radio transients without the need for external triggers, enabling radio-selected populations to be compared with those from other wavelengths. 

Here we report on a remarkable radio transient discovered by jointly searching the first half of the first epoch of VLASS (VLASS\,1.1) and the VLA Faint Images of the Radio Sky at Twenty centimeters (FIRST) survey \citep{FIRST}.\footnote{A similar search yielded the discovery of the luminous extragalactic radio transient FIRST\,J141918.9$+$394036 \citep{j1419}.} The FIRST survey was conducted at a frequency of 1.4\,GHz, and covered $\sim10,000$\,deg$^{2}$ of the northern sky mostly between 1994 and 1999 with an angular resolution of 5\arcsec and a continuum-image rms of $150\,\mu$Jy per beam. We discovered a source (see \S\ref{sec:sample}), FIRST\,J153350.8$+$272729 (hereafter J1533$+$2727), which was detected in FIRST in 1995 with a flux density of 9.7\,mJy, but was undetected in VLASS in 2017.
Further searches of archival radio data revealed that J1533$+$2727 was detected in 1986 and 1987 by the Green Bank 300-foot telescope \citep{gb6} at 4.85\,GHz, with a mean flux density of 51\,mJy. 
We performed multi-band observations with the VLA on 2019 May 14 that re-detected the source at a level consistent with the VLASS upper limit (\S\ref{sec:archival}), notably with a flux density of just $132\,\mu$Jy at 5\,GHz. 
J1533$+$2727 is associated with the nucleus of a galaxy (SDSS\,J153350.89$+$272729) at a (luminosity) distance of 147\,Mpc (see \S\ref{sec:host}). 
The fading of J1533$+$2727 by nearly a factor of 400 over 33 years at $\sim5$\,GHz is strong evidence for its transient nature, and the preferred interpretation for its origin is of a TDE (see \S\ref{sec:Interpretation}). 

Although nearly a hundred TDE candidates are now cataloged\footnote{See http://tde.space.}, only $\sim10\%$ of TDEs exhibit radio emission \citep{tde_sample,marin_tde}. The three most distant of the radio TDEs (Sw\,J1644$+$57, Sw\,J2058$+$05, and Sw\,J1112-82 at redshifts of 0.35, 1.18, and 0.89 respectively) were first discovered through transient $\gamma$-ray emission corresponding to the launch of a nascent relativistic jet. The remaining radio TDEs are powered by mildly relativistic outflows that drive shocks into the circum-nuclear medium, and peak at radio luminosities that are two to three orders of magnitude below the three relativistic radio TDEs. Only one TDE candidate, CNSS\,J0019$+$00 \citep{marin_tde}, has previously been first discovered through its radio emission. In general, the radio emission generated by extragalactic explosions (e.g., supernovae, $\gamma$-ray bursts, and TDEs) is enhanced in the presence of more energetic outflows, and denser circum-explosion material. TDEs discovered through their radio emission, rather than through X-ray or optical emission associated with accreting material, may offer a novel selection of TDE phenomena and host galaxies.


Throughout this work, we adopt the following cosmological parameters: $H_{0}=67.7$\,km\,s$^{-1}$\,Mpc$^{-1}$, $\Omega_{M}=0.3089$, and $\Omega_{\Lambda}=0.6911$ \citep{planck15}. 


\section{Discovery of J1533+2727}\label{sec:sample}

Two independent efforts discovered J1533$+$2727 by comparing catalogs of sources from VLASS\,1.1 and FIRST. Each effort first generated source catalogs from the VLASS\,1.1 quick-look images\footnote{https://archive-new.nrao.edu/vlass/quicklook/}, using either the Aegean \citep{aegean1,aegean2} or PyBDSF \citep{pybdsf} source finding software. Details of how the source finding algorithms were applied will be presented in future works that describe larger samples of transients (e.g., Dong et al., in prep.). 
Once the VLASS\,1.1 source catalogs were made, we performed cross-matches between unresolved sources in the VLASS\,1.1 and FIRST catalogs, with a specific focus on finding FIRST sources not present in VLASS\,1.1. 
In one of our efforts we only considered sources with a more than $75\%$ decrease in measured flux densities between FIRST and VLASS. In the other effort we only considered sources 
detected at $>2.3$\,mJy in FIRST and undetected in VLASS (i.e., with 3\,GHz flux densities $<0.5$\,mJy), which were additionally coincident with the nuclei of spectroscopically detected galaxies in SDSS\,DR14 \citep{sdss14} at redshifts $z<0.1$. J1533+2727 was noteworthy as one of the brightest sources to pass all our selection thresholds. The position of this source in the FIRST catalog is (R.A. J2000, decl. J2000) = (15:33:50.884, $+$27:27:29.57), with uncertainties of 0.4\arcsec~in each coordinate \citep{FIRST}. 

\begin{deluxetable*}{cccc}
\tabletypesize{\footnotesize}
\tablewidth{0pt}
\centering
\tablecaption{ Radio Observations of FIRST\,J153350.8$+$272729. \label{tab:sample}}
\tablehead{
\colhead{Epoch} & \colhead{Survey} & \colhead{Frequency (GHz)} & \colhead{Flux density (mJy)} 
}
\startdata 
1968 & Bologna & 0.408 & $< 750$ \\
1974 -- 1983 & Texas & 0.365 & $< 1200$ \\
1983 April 2 -- 21 & GBNSS & 1.4 & $< 300$ \\ 
1986 November 6 -- December 13 & GB6 & 4.85 & $65\pm8$ \\
1987 September 30 -- November 1 & GB6 & 4.85 & $42\pm8$ \\
1987 January & GEETEE & 0.0345 & $<$ 15000   \\
1995 April 16 & NVSS & 1.4 & $9.1\pm0.5$ \\
1995 November 65 & FIRST & 1.4 & $9.7\pm0.1$ \\
1998 May 22 & CLASS & 8.46 & $<1.5$ \\
2001 September 14 & CLASS & 1.425 & $<3$ \\
2001 September 14 & CLASS & 4.86 & $<1.5$ \\
2001 September 14 & CLASS & 8.46 & $<1.5$ \\
2001 September 14 & CLASS & 22.46 & $<7.5$ \\
2006 & VLSSr & 0.074 & $< 300$\\
2010 April -- 2012 March & TGSS &0.15 & $< 15$\\
2017 October 2 & VLASS & 3 & $<0.46$ \\
2019 May 14 & 19A-470 & 1.52 & $0.36\pm0.03$ \\
2019 May 14 & 19A-470 & 5 & $0.132\pm0.009$ \\
2019 May 14 & 19A-470 & 7 & $0.100\pm0.009$ \\
\enddata
\tablecomments{All upper limits are at the $3\sigma$ level.
References: Bologna Sky Survey \citep[Bologna;][]{BOLOGNA}, Texas Survey of Radio Sources at 365 MHz \citep[Texas;][]{TEXAS}, Green Bank Northern Sky Survey \citep[GBNSS;][]{GBNSS}, Green Bank 6 cm survey \citep[GB6;][]{gb6cat}, Gauribidanur Telescope \citep[GEETEE;][]{GTEE}, Faint Images of the Radio Sky at Twenty centimeters \citep[FIRST;][]{FIRST}, NRAO VLA Sky Survey \citep[NVSS;][]{nvss}, Cosmic Lens All-Sky Survey \citep[CLASS;][]{class}, VLA Low-frequency Sky Survey \citep[VLSSr;][]{VLSSr}, TIFR GMRT Sky Survey Alternative Data Release \citep[TGSS;][]{TGSS}, VLA Sky Survey \citep[VLASS;][]{VLASS}.}
\end{deluxetable*}

\section{Archival and follow-up radio observations}\label{sec:archival}

We searched a selection of existing radio-survey catalogs and data sets for detections of J1533+2727. The results are summarized in Table~\ref{tab:sample}. J1533+2727 is cataloged in the FIRST survey with a flux density of $9.7\pm0.1$\,mJy, on an observing epoch of 
1995 November 06. The source is also present in the NVSS catalog with a flux density of $9.1\pm0.5$\,mJy on 1995 April 16. Although the formal $3\sigma$ upper limit on the flux density of J1533+2727 in the VLASS quick-look images is 0.38\,mJy (based on the per-pixel rms at the source location), we adopt an upper limit of 0.46\,mJy to account for errors in the flux scale \citep{pb17}. The VLASS observation epoch was 2017 October 02. We next searched the VLA archive\footnote{\url{https://archive.nrao.edu/archive/advquery.jsp}} for data obtained at the position of J1533+2727, and found that this source had been observed (in a targeted observation, at the center of the primary beam) by the Cosmic Lens All-Sky Survey \citep[CLASS;][]{class} on 1998 May 22 at 8.46\,GHz (VLA project AM0593), and by a wideband survey of sources with rising spectra between 1.4\,GHz and 4.8\,GHz on 2001 September 14 (VLA project AG0617). We re-analyzed these data using standard tasks from the Common Astronomy Software Applications \citep[CASA, version 5.1.1;][]{casa}, and finding no source at the position of J1533+2727 derived the upper limits on its flux density reported in Table~\ref{tab:sample}. As above, these upper limits were derived using the per-pixel rms at the source location. 

The selection of J1533+2727 as a CLASS source implied that J1533+2727 is present in the Green Bank 300-foot telescope 6\,cm survey catalog \citep[GB6;][]{gb6cat} with a flux density in excess of 30\,mJy \citep{class} at 4.85\,GHz. Indeed, the GB6 catalog lists a $51\pm6$\,mJy source (GB6\,J1533$+$2728) at a position of (R.A. J2000, decl. J2000) = (15:33:49.9$\pm$0.8, $+$27:28:12$\pm$12), where a substantial component of the error is due to pointing errors of order 8\arcsec~\citep{gb6cat}. Despite the 46.8\arcsec~offset between the FIRST and GB6 sources, the association is considered likely as the next closest source in FIRST or VLASS to the GB6 position is offset by 379\arcsec, which is greater than the 3.5\,arcmin full-width half-maximum of the GB6 survey beam.\footnote{The final CLASS sample was chosen by associating GB6 sources with NVSS sources within a 70\arcsec~separation cut, which is explained by the much greater positional uncertainty of NVSS than FIRST.} The GB6 survey catalog is actually comprised of observations obtained over two epochs, between 1986 November 6 and December 13, and between 1987 September 30 and December 1.\footnote{Survey operations in late 1988 with the 300-foot telescope were increasingly affected by pointing errors that rendered the data unusable.} Single epoch maps were converted into source catalogs by \citet{gb6var}\footnote{Currently available at  \url{https://phas.ubc.ca/~gregory/RadioAstronomy.html}.}. The 1986 observations contain J1533+2727 with a flux density of $65\pm8$\,mJy, and the 1987 observations showed a flux density of $42\pm8$\,mJy. 

The remarkable fading of J1533+2727 over 31 years between 1986 and 2017 motivated follow-up VLA observations (VLA program 19A-470). We obtained data in the B configuration (antenna separations between 0.21\,km and 11.1\,km) on 2019 May 14 in the L (1--2\,GHz) and C (4--8\,GHz) bands, using standard continuum observing setups and CASA data-reduction procedures. The absolute flux-scale and bandpass calibrator was 3C286, and time-variable complex gain calibration was accomplished using J1513+2338. No self-calibration was conducted. We detected J1533+2727 in both bands at a position consistent with the FIRST position within 0.2\arcsec. The measured flux densities were $0.36\pm0.03$\,mJy at 1.52\,GHz, $0.132\pm0.009$\,mJy at 5\,GHz, and $0.100\pm0.009$\,mJy at 7\,GHz. The measurements are consistent with a single power law (flux density $S(\nu)\propto\nu^{\alpha}$, where $\nu$ is the frequency) with spectral index $\alpha=-0.840\pm0.003$. These measurements do not include a 3--5\% uncertainty in the VLA flux-density scale \citep{pb17}.

\section{Host-galaxy properties}\label{sec:host}

\begin{figure*}[ht]
  \centering
    \includegraphics[width=0.35\textwidth]{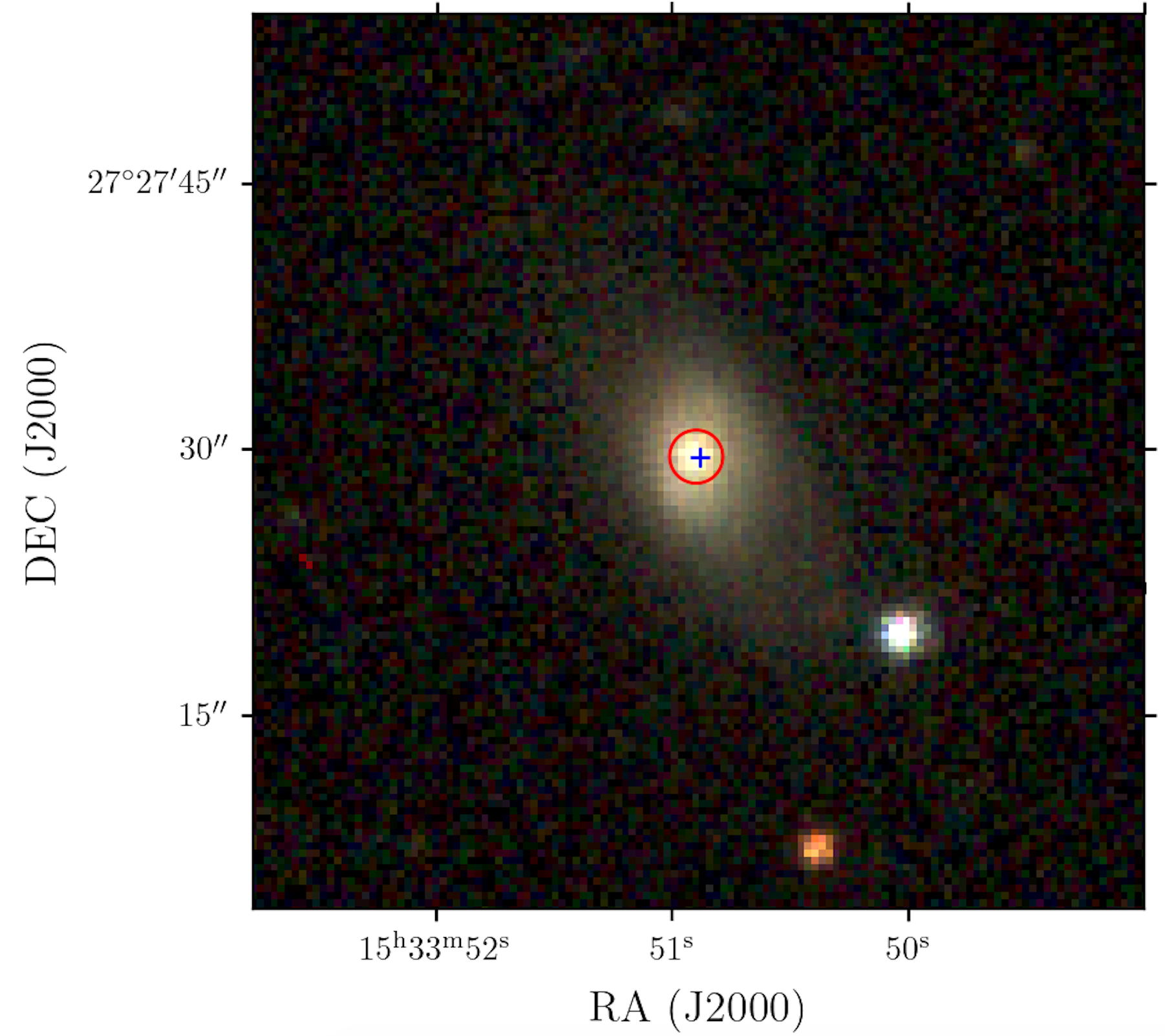}
    \hspace{0.6cm}
    \includegraphics[width=0.5\textwidth]{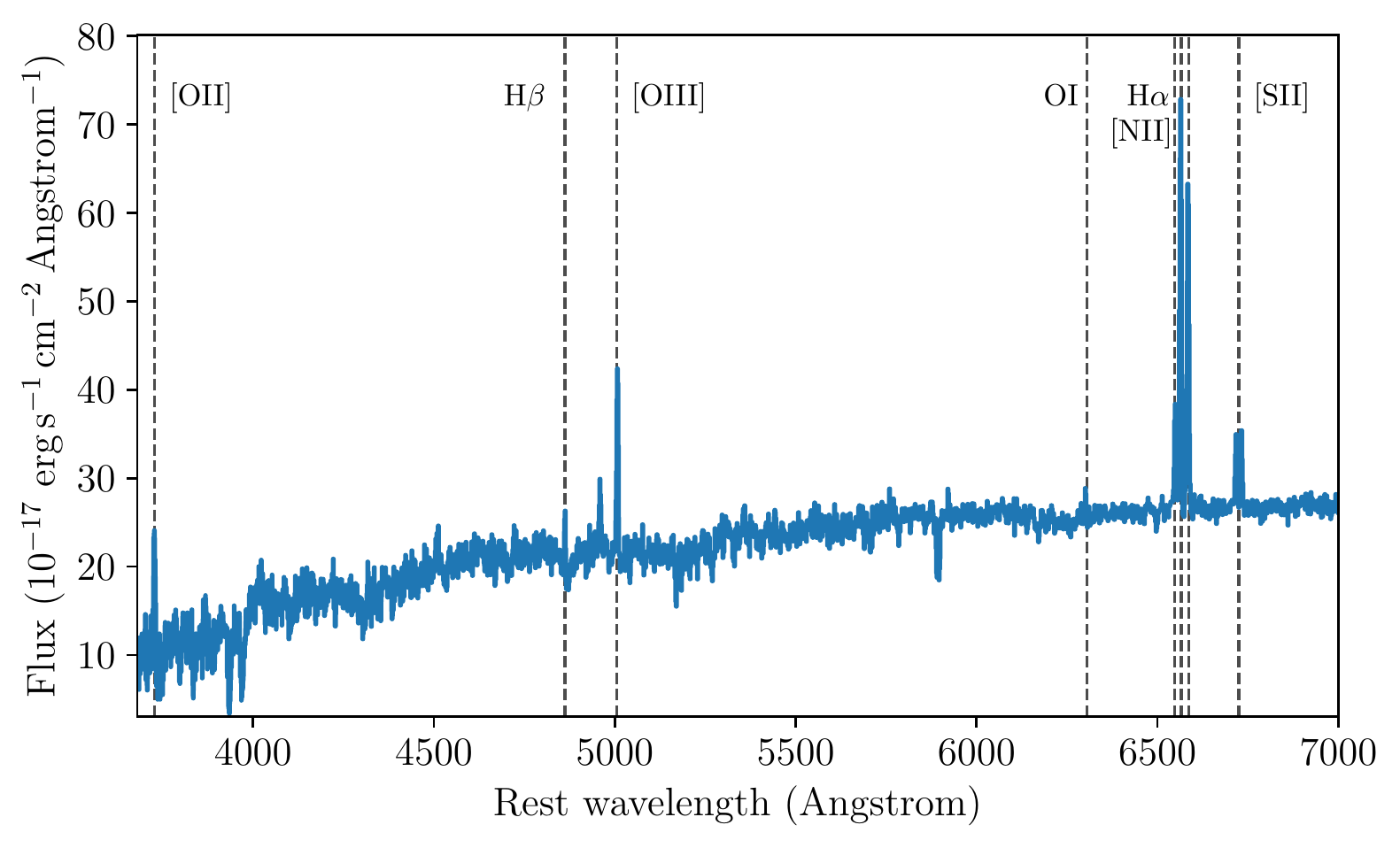}
  \caption{{\em Left:} Three-color composite in the SDSS $g$, $r$ and $i$ bands of the host galaxy of J1533+2727, SDSS\,J153350.89+272729. The radio position of J1533+2727 is shown as a blue cross, and the spatial extent of the SDSS fiber input on the sky is shown as a red circle \citep{sdss_spec}. {\em Right:} SDSS spectrum \citep{sdss16} of SDSS\,J153350.89+272729 obtained on 2007 March 21. Some relevant emission lines are labeled, which are indicative of weak Type II Seyfert activity.}
  \label{fig:host}
\end{figure*}

The centroid of the FIRST position of J1533+2727 is located 0.150\arcsec~from the optical center of the galaxy SDSS\,J153350.89+272729. With a FIRST positional uncertainty of 0.400\arcsec, J1533+2727 is therefore consistent with being coincident with the galaxy nucleus. Figure~\ref{fig:host} shows the SDSS DR16 \citep{sdss16} image and spectrum of this galaxy, which we adopt as the host of J1533+2727. The host galaxy lies at a redshift of $z=0.03243\pm0.00001$ (luminosity distance of $147.14\pm0.01$\,Mpc), and the difference between the radio position of J1533+2727 and the center of light of the host galaxy corresponds to a projected separation of just $107$\,pc. 
An inspection of the SDSS optical spectrum indicates weak Type II Seyfert activity, according to standard line-ratio diagnostics \citep{kewley}, with $\log{\rm ([NII]/H\alpha)}=-0.05$ and $\log{\rm ([OIII]/H\beta)}=0.78$. Stellar population synthesis fits to the SDSS photometry indicate a stellar mass of between $10^{10.45}M_{\odot}$ and $10^{10.49}M_{\odot}$, and an ongoing star-formation rate of $\sim0.6\,M_{\odot}$\,yr$^{-1}$ \citep[`stellarMassFSPSGranEarlyDust' table;][]{sdssspec}. The quoted uncertainties are purely statistical, and do not reflect the range of possible parameterizations. The galaxy was classified morphologically as a spiral by \citet{class_SPIRAL}. 

The absolute rest frame magnitudes of the bulge in the $g$ and $r$ bands were calculated by \citet{Abs_Mag} using Galaxy IMage 2D \citep{GIM2D}, assuming extinction values obtained from the SDSS database. We transformed these bulge absolute magnitudes, $M_{\rm g,bulge}=-18.52$ and $M_{\rm r,bulge}=-19.39$ to the $V$ band, $M_{\rm V,bulge}=-19.04$, according to formulas from \citet{JESTER}. We then applied the relation between SMBH mass and bulge luminosity from \cite{Lum_to_mass} to estimate the total black hole mass to be $\log M_{BH}/M_{\odot}=7.6 \pm 0.2$.

\section{Discussion}\label{sec:Interpretation}

\subsection{The nature of J1533+2727}

We now interpret these observations in terms of three hypotheses for J1533+2727:
\begin{itemize}
    \item AGN variability.
    \item An engine-driven transient associated with a supernova.\footnote{We do not consider standard supernovae because the peak radio luminosity of  J1533+2727 ($1.6\times10^{30}$\,erg\,s$^{-1}$\,Hz$^{-1}$) is a factor of eight greater than even the most luminous radio supernova \citep[PTF\,11qcj;][]{11qcj}.}
    \item A jet or outflow powered by a TDE. 
\end{itemize}

We first augment the observations presented above with archival ROSAT/PSPC pointings that included J1533+2727 in the field of view. We highlight two observations in particular: a 415\,s pointing on MJD~48102 (1990 July 30; sequence id rs931238n00; around three years after the last Green Bank detection) obtained as part of the ROSAT All-Sky Survey, and a 13932\,s pointing a year later on MJD~48449 (1991 July 12; sequence id RP201103N00) obtained as part of a long exposure on $\alpha$ Cor Bor. We used the \texttt{sosta} tool in the \texttt{XIMAGE} package, and the exposure maps associated with the observations, to derive $3\sigma$ upper limits on the count rates at the position of J1533+2727. These were 0.125\,cts\,s$^{-1}$ and 0.0108\,cts\,s$^{-1}$ on the respective dates (in the standard PSPC 0.1--2.4\,keV band). We then converted these to upper limits on the 2--10\,keV luminosity assuming a photon index of 2, and a Galactic neutral-hydrogen column density of $n_{H}=2.9\times10^{20}$\,cm$^{-2}$ derived from the HI4PI neutral hydrogen column map \citep{hi4pi}. These upper limits were $L_{X}<3.0\times10^{42}$\,erg\,s$^{-1}$ and $L_{X}<2.6\times10^{41}$\,erg\,s$^{-1}$ on the respective dates. 

These X-ray upper limits are low in comparison with the radio luminosity of J1533+2727, if J1533+2727 represents an active black hole. The black hole fundamental plane relates the mass, 5\,GHz radio luminosity, and 2--10\,keV X-ray luminosity of actively accreting objects across nine orders of magnitude in black hole mass, and hints at a fundamental link between accretion rate and jet power. We used the latest iteration of this relation \citep{gkc+19} to derive predicted upper limits on the 5\,GHz radio luminosity of J1533+2727 at the epochs of the X-ray observations. Given the derived SMBH mass, by assuming Poisson statistics for the X-ray upper limits, and by using a Monte Carlo technique to account for the uncertainty and intrinsic scatter in the black hole fundamental plane, we calculated 95\% confidence upper limits on the expected radio luminosity of $4.8\times10^{38}$\,erg\,s$^{-1}$ and $9.8\times10^{37}$\,erg\,s$^{-1}$ corresponding to the two X-ray observations. These in turn imply radio flux-density upper limits of 3.7\,mJy and 0.8\,mJy respectively, where we divided the luminosities by the frequency to derive the spectral luminosities following \citet{gkc+19}. If J1533+2727 was indeed this faint during the X-ray observations, a much more rapid evolution is implied between the Green Bank 4.85\,GHz detection and these epochs than at later times. Furthermore, unless the source re-brightened between the X-ray observations and the NVSS and FIRST detections, an unrealistic spectral index steeper than $-2$ is implied between 1.4\,GHz and 5\,GHz for the FIRST detection. We conclude that J1533+2727 was inconsistent with the black hole fundamental plane when the ROSAT X-ray observations were undertaken. 

This suggests that J1533+2727 was not actively accreting as an AGN at this time, which is in tension with the hypothesis of ongoing AGN variability. This is however not in tension with the TDE hypothesis, because the accretion could have ceased \citep[e.g.,][]{levan15}. When detected in the Green Bank survey, J1533+2727 was also more luminous at a wavelength of 6\,cm than the cores of any of the 52 nearby Seyfert galaxies observed by \citet{hu01}, besides Perseus~A (NGC\,1275; Seyfert~1.5) and NGC\,1167 (Seyfert~II; beyond the magnitude-completeness limit of the survey). Additionally, J1533+2727 is likely more variable than any of the 12 Seyferts observed by \citet{mfn+09}, within whose sample the maximum variability in seven years at 8.4\,GHz was a factor of three. We therefore proceed to consider hypotheses (2) and (3) given above. 

Some insight can be gained by analyzing J1533+2727 as a synchrotron transient, despite the lack of detailed spectral information. That J1533+2727 represents synchrotron emission is evident given the brightness temperature ($\gtrsim5.3\times10^{9}$\,K) implied by the extreme variability of the source.\footnote{A useful working definition of a variability timescale that can be used to calculate a light-crossing time is a timescale over which the modulation index, defined as the variability range divided by the mean source flux density, is greater than unity. Adopting a timescale of six years (between the FIRST/NVSS observations and L-band VLA observations in 2001), we find a brightness temperature in excess of $5.3\times10^{9}$\,K.}
We can also derive rough constraints on the source radius, $R$, the energy required to power the source, $E$, and the electron number density, $n_{e}$, of the medium into which the source is expanding. The constraints are based on a fiducial $65$\,mJy maximum flux density measured at $4.85$\,GHz, and the assumption that the optically thin spectral index observed in our 2019 observations of the source of $\alpha=-0.84$ is representative of a non-evolving relativistic electron energy distribution $N(E)\propto E^{-p}$ with $p=-2\alpha+1=2.68$. In the following, we assume (\textit{a}) equipartition between the energy in relativistic electrons and in magnetic fields within the source, (\textit{b}) that the source is expanding sub-relativistically, (\textit{c}) that the source is spherically symmetric with a filling factor of unity, and (\textit{d}) that the relativistic electrons are accelerated in a strong (forward) shock that deposits 10\% of its energy in the electrons, and 10\% of its energy in magnetic fields (i.e., $\epsilon_{e}=\epsilon_{B}=0.1$ in usual terms).\footnote{In our calculation, we adopt $c_{5}=9.68\times10^{-24}$ and $c_{6}=8.10\times10^{-41}$ from \citet{Pacholczyk1970}.} The non-relativistic assumption further implies that the spectral peak was associated with synchrotron self-absorption rather than the minimum relativistic-electron energy \citep[e.g.,][]{c98}. 

In this scenario, following \citet{c98}, $R\propto S_{p}^{(p+6)/(2p+13)}\nu_{p}^{-1}$, and $E\propto S_{p}^{(3p+14)/(2p+13)}\nu_{p}^{-1}$, where $\nu_{p}$ is the peak frequency and $S_{p}$ is the peak flux density. If we are simply constraining the values of $R$ and $E$ when $\nu_{p}=4.85$\,GHz, we are setting lower limits on both quantities. This is also essentially the case if we are constraining the values of $R$ and $E$ during the 1986 Green Bank observations.\footnote{If the spectrum was optically thick and $\nu_{p}>\nu=4.85$\,GHz, $S_{p}\propto(\nu_{p}/\nu)^{2}$ implies a nearly fixed estimate of $R$ regardless of the true value of $\nu_{p}$, and a larger value of $E$. If the spectrum was optically thin, $R$ and $E$ would clearly both be larger.} The electron number density is derived by applying the Rankine-Hugoniot jump conditions in the strong shock limit \citep[Equation~(15) of][]{18cow}, with the further assumption of the shock velocity being given by the source size divided by its lifetime, $T$. In this case, $n_{e}\propto S_{p}^{-(2p+16)/(2p+13)}\nu_{p}^{4}$. In summary, using the 1986 measurement, we find 
\begin{align}
R &=\begin{aligned}& 2.3\times10^{17} \left(\frac{S_{p}}{\rm 65\,mJy}\right)^{0.47}\left(\frac{\nu_{p}}{\rm 4.85\,GHz}\right)^{-1}\,{\rm cm}\end{aligned} \\[\jot]
E &=\begin{aligned}& 7.5\times10^{50} \left(\frac{S_{p}}{\rm 65\,mJy}\right)^{1.20}\left(\frac{\nu_{p}}{\rm 4.85\,GHz}\right)^{-1}\,{\rm erg}\end{aligned} \\[\jot]
n_{e} &=\begin{aligned}& 1.6\times10^{-3} \left(\frac{S_{p}}{\rm 65\,mJy}\right)^{-1.16}\left(\frac{T}{\rm 1\,day}\right)^{2} \\ 
&\left(\frac{\nu_{p}}{\rm 4.85\,GHz}\right)^{4}\,{\rm cm}^{-3}.\end{aligned} 
\end{align}
These equations, including the constants of proportionality, directly reproduce Equations (8), (12), and (16) of \citet{18cow}, for our value of $p$. Although the radius estimate is only mildly sensitive to the assumptions above, a departure from equipartition like that observed for Sw\,J1644$+$57 \citep{tarraneh}, where $\epsilon_{B}=0.001$ was inferred assuming $\epsilon_{e}=0.1$, would increase the energy estimate by two orders of magnitude. 


\begin{figure*}[!ht]
  \centering
    \includegraphics[width=\textwidth]{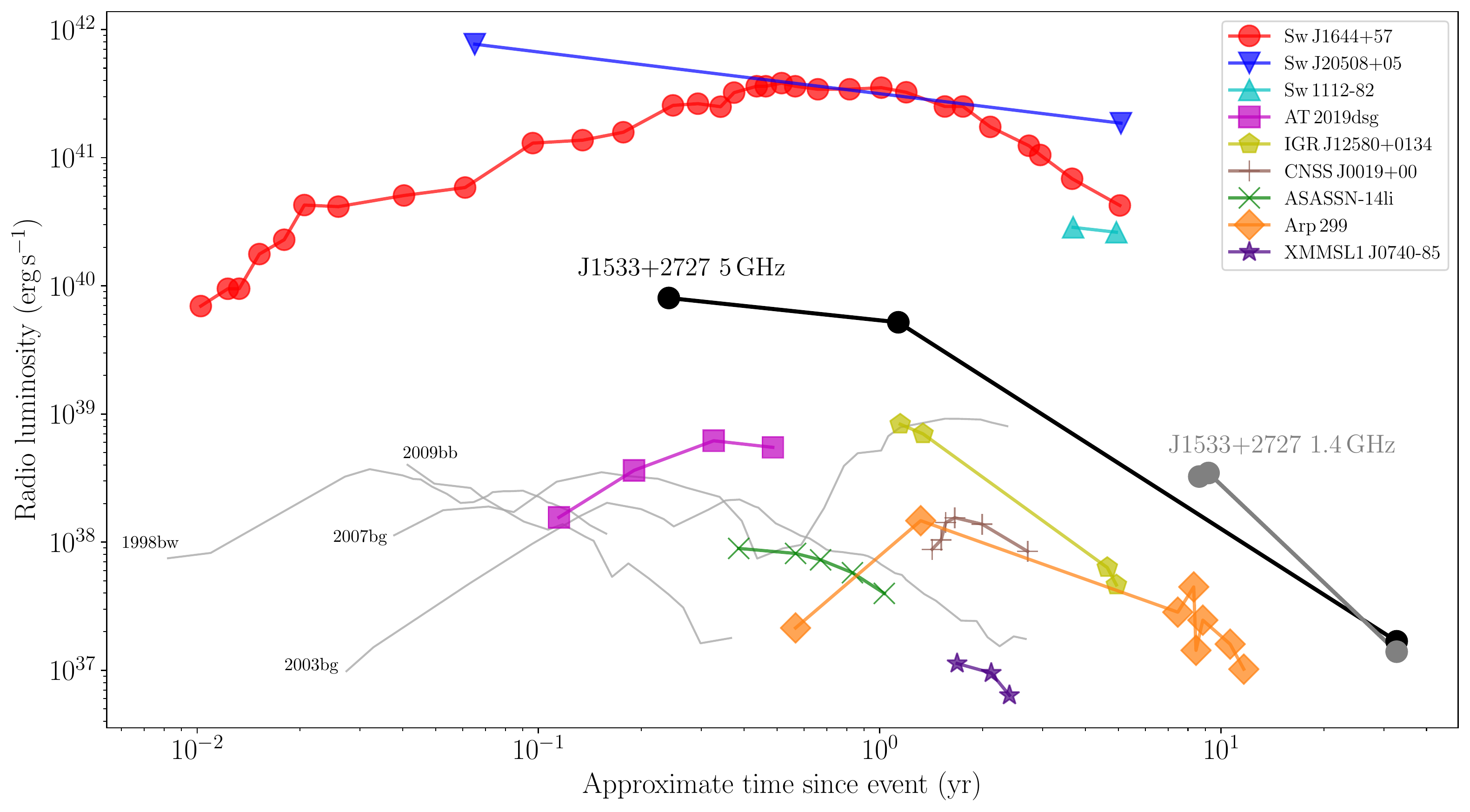}
  \caption{Lightcurves of all TDEs detected at radio wavelengths. All TDE data are at 5\,GHz, except for the 1.4\,GHz data on J1533$+$2727, and 8.4\,GHz data on AT2019dsg and Arp\,299. Data were collated from \citet{tde_sample} and references therein. The time since the event for the first data point on J1533$+$2727 (88 days) was derived assuming expansion at the speed of light to that time (see text for details). Also shown are representative radio lightcurves of four luminous supernovae: the relativistic-outflow events 1998bw \citep[4.8\,GHz;][]{1998bw} and 2009bb \citep[8.46\,GHz;][]{2009bb}, and the energetic Type Ic broad line events 2003bg \citep[8.5\,GHz;][]{2003bg} and 2007bg \citep[8.5\,GHz;][]{2007bg}. In the latter two supernovae the late-time radio emission was enhanced by a dense, structured circumstellar medium.}
  \label{fig:lightcurve}
\end{figure*}

These results provide evidence that J1533+2727 represents a relativistic jet/outflow from a TDE. First, the lower limit on the energy in the outflow is greater than that of any known stellar cataclysm \citep[e.g.,][]{229B_m} besides classical, on-axis long gamma-ray bursts (LGRBs).\footnote{Radio calorimetry of the ejecta of cosmic explosions traces the fastest ejecta, and therefore cannot be directly related to the total energies in outflows with a range of velocities, like supernovae \citep[e.g.,][]{berger03}.} Additionally, most LGRBs have apparent expansion velocities of $\Gamma\beta\gtrsim3$ (here, $\Gamma$ is the bulk Lorentz factor of the emitting material, and $\beta=v/c$ is the normalized expansion velocity). However for J1533+2727, following \citet{sari98}, the characteristic frequency corresponding to radiation from the lowest-energy relativistic electrons, $\nu_{m}$, was likely lower than 4.85\,GHz in 1986, because the source declines between 1986 and 1987. This frequency is related to $E$ and $T$ in the case of adiabatic evolution, which implies a lifetime:
\begin{equation}
    T\gtrsim22\left(\frac{E}{7.5\times10^{50}\,{\rm erg}}\right)^{1/2}\,{\rm days}.
\end{equation}
This in turn implies $\Gamma\beta\lesssim4$, and $n_{e}\lesssim 0.8\,{\rm cm}^{-3}$. Although this makes the LGRB scenario somewhat fine-tuned, the derived parameters are consistent with some GRBs \citep[e.g., GRB\,980703;][]{perley17}. Indeed, GRB\,980703 exhibits late-time emission 16 years post-burst that is similar to J1533+2727 \citep{perley17}. However, the projected offset between J1533+2727 and the center of light of its host galaxy of $\approx 100$\,pc is inconsistent with more than 95\% of the LGRB population \citep{grboffset}. Additionally, the high stellar mass and lack of evidence for an ongoing starburst in the host galaxy of J1533+2727 are inconsistent with typical LGRB hosts \citep{taggart}. We therefore favor the TDE scenario.

A comparison between the radio lightcurve of J1533+2727 and the remainder of the TDE population is shown in Figure~\ref{fig:lightcurve}. The post-explosion time  of the 1986 epoch (88 days) was derived assuming a nominal expansion velocity of $c$, which would imply $n_{e}\sim12$\,cm$^{-3}$. Please note however that sub-relativistic expansion was assumed to derive the constraints above, and this is therefore for illustrative purposes only. The high radio luminosity and outflow energy relative to several radio-detected TDEs is suggestive of a relativistic jet, rather than a wide-angle outflow \citep{tde_sample}. We note that if the emission region were significantly aspherical, with a non-unity filling factor, some of the above conclusions would be altered by factors of a few \citep{duran}.





\subsection{Implications for the TDE population}

We have established that J1533+2727 is a remarkable radio transient and a likely TDE afterglow. Using the FIRST survey, and assuming the detection of just one such source, we can calculate a lower limit on the occurrence of sources like J1533+2727. The 1\,mJy minimum flux density of the FIRST catalog and the peak luminosity of J1533+2727 in FIRST implies a detectable distance of 452\,Mpc. The FIRST and VLASS\,1.1 sky surveys have an overlapping sky coverage of $\sim6000$\,deg$^{2}$, and thus the detectable distance corresponds to an observable volume of 0.056\,Gpc$^3$. J1533+2727 emitted above its detected FIRST luminosity for at least 8 years between the Green Bank and FIRST detections. We can therefore infer a lower limit on the volumetric rate of approximately 2.2\,Gpc$^{-3}$yr$^{-1}$, or $\sim1$\% of the observed TDE rate \citep{rate}. We do not take this analysis further because the search for TDEs detected in FIRST is ongoing. 

\begin{figure}[!ht]
  \centering
    \includegraphics[width=0.45\textwidth]{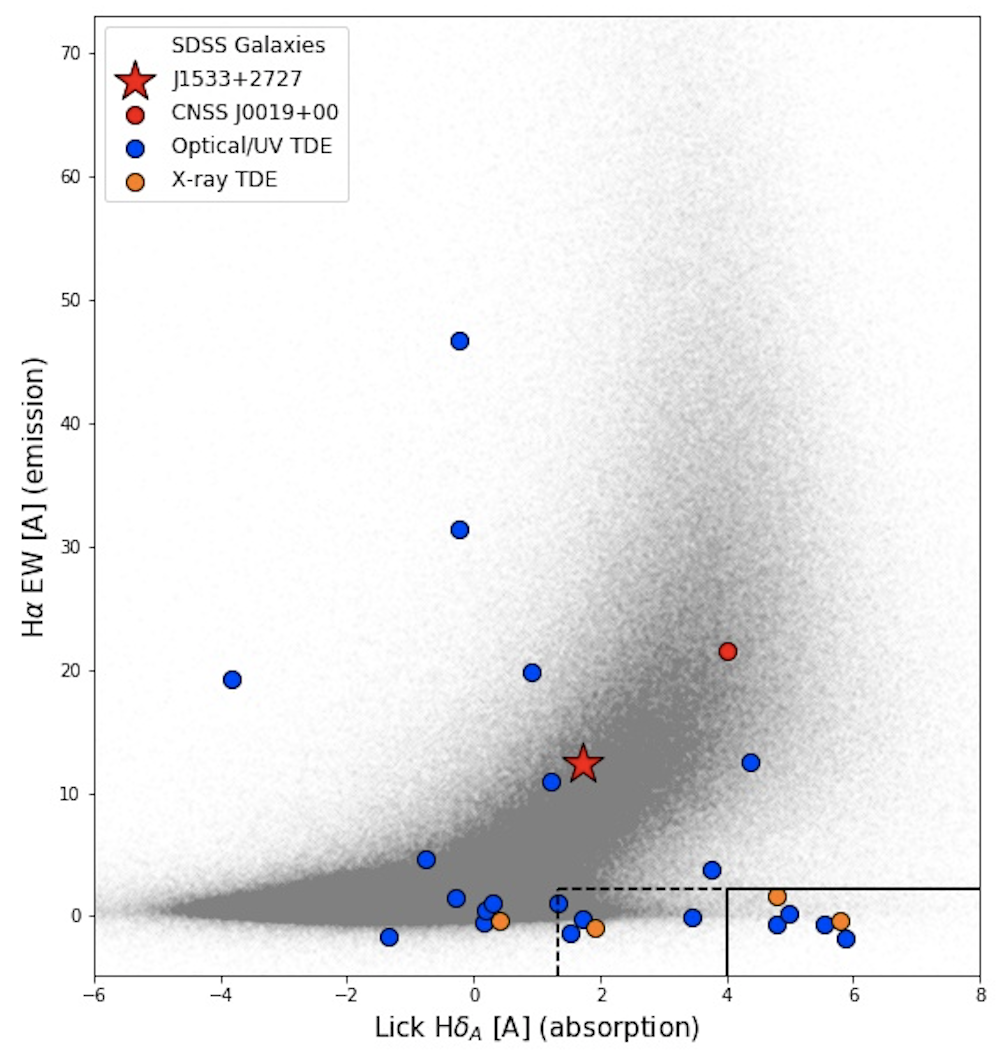}
  \caption{Plot adapted from \citet{french_host} (their Figure~5 -- see \citet{french_host} and \citet{decker_2016} for details) showing spectral indices of a sample of SDSS galaxies, and selected TDE hosts including J1533+2727. H$\alpha$ emission traces current star formation while H$\delta$ absorption traces star-formation activity in the past $\sim$Gyr. Post-starburst / quiescent Balmer-strong galaxies comprising 0.2\% (solid box) and 2\% (dashed box) of the parent SDSS sample are at the lower right of the plot. The hosts of optically and X-ray selected TDEs are over-represented among post-starburst galaxies. Although only two radio-selected TDE candidates have been identified so far (J1533+2727 and CNSS\,J0019+00), neither is hosted by a post-starburst galaxy.}
  \label{fig:host_distribution}
\end{figure}

Our results add to the emerging picture of the diversity of TDE-driven jets/outflows from supermassive black holes. Although it has long been known that $\sim50\%$ of the mass of a disrupted star is likely to be unbound \citep[e.g.,][]{rees1988}, the geometry and kinematics of such outflows are poorly constrained, as are any jets/outflows powered by the accretion of the remaining 50\% of the mass.  The radio luminosity and derived outflow energy of J1533+2727 fills the gap between the three relativistic TDEs identified through their prompt high-energy emission, and the remaining TDE sample (Figure~\ref{fig:lightcurve}). The host galaxy of J1533+2727 and its central supermassive black hole appear otherwise unremarkable with respect to the TDE population \citep{french_host,BH_mass}. Although the black hole mass is somewhat high relative to optically selected TDEs \citep{BH_mass}, stars with a wide range of masses ($\gtrsim0.3M_{\odot}$) are expected to be disrupted by such black holes \citep{kochanek}. The optical spectrum of the host of J1533+2727 shows emission lines characteristic of the narrow-line regions of Type II Seyferts; this nuclear activity must have been ongoing prior to the transient event, given the large expected sizes of narrow-line regions \citep[e.g.][]{bennert06}. This is similar to the host of the radio-discovered TDE CNSS\,J0019$+$00 \citep{marin_tde}. The presence of nuclear activity in the spectrum of the J1533+2727 host makes it difficult to determine whether or not it is a post-starburst galaxy, although we note that TDEs are found to be over-represented in galaxies that are evidently post-starburst from their optical spectra \citep[Figure~\ref{fig:host_distribution};][]{decker_2016,french_host}.
We speculate that TDEs discovered in radio transient surveys will have substantially different selection effects, especially with regards to AGN activity and extinction, than the selection effects present in optical and soft X-ray surveys that dominate TDE discoveries today. 

\section{Conclusions}\label{sec:Conclusions}

We present the discovery of the candidate TDE FIRST\,J153350.8$+$272729 using the GB6, FIRST, and VLASS radio surveys. The source was first detected in 1986 with a flux density of 65\,mJy at 4.85\,GHz, and has been monotonically fading ever since. This is the second TDE candidate to be solely identified at radio wavelengths. Its radio luminosity (observed maximum of $8\times10^{39}$\,erg\,s$^{-1}$), and the implied energy in the outflow generated by the TDE ($\gtrsim7\times10^{50}$\,erg), fill a gap between most radio-detected TDEs and the three high-redshift events that were first discovered through their prompt $\gamma$-ray emission. Little more can be said about the nature of the outflow and the medium into which it propagates, because we have only observed the optically thin component of the radio spectral energy distribution. The host galaxy, at a distance of 147\,Mpc, is largely unremarkable (inferred supermassive black hole mass of $4\times10^{7}M_{\odot}$), and shows signatures of Type II Seyfert activity. We anticipate that ongoing surveys for radio transients like VLASS, and with the Australian Square Kilometre Array Pathfinder \citep{vast}, and the Aperture Tile In Focus at the Westerbork Synthesis Radio Telescope \citep{apertif}, together with dedicated radio follow-up observations will yield several such TDEs that may help untangle selection effects in surveys in other bands.  

\acknowledgements

Contributions from GZ and JC were made through the Harvard Science Research Mentoring Program \citep[SRMP;][]{graur2018}. Support for this program is provided by the National Science Foundation under award AST-1602595, City of Cambridge, the John G. Wolbach Library, Cambridge Rotary, and generous individuals. We thank Or Graur for useful discussions on the science, and for coordinating the 2018--2019 SRMP that made this research possible. The Dunlap Institute is funded through an endowment established by the David Dunlap family and the University of Toronto. H.D. and B.M.G. acknowledge the support of the Natural Sciences and Engineering Research Council of Canada (NSERC) through grant RGPIN-2015-05948, and of the Canada Research Chairs program. M.R.D. acknowledges support from the NSERC through grant RGPIN-2019-06186, the Canada Research Chairs Program, the Canadian Institute for Advanced Research (CIFAR), and the Dunlap Institute at the University of Toronto. C.J.L. acknowledges support under NSF grant 2022546. The National Radio Astronomy Observatory is a facility of the National Science Foundation operated under cooperative agreement by Associated Universities, Inc. The Pan-STARRS1 Surveys (PS1) and the PS1 public science archive have been made possible through contributions by the Institute for Astronomy, the University of Hawaii, and others. Funding for the Sloan Digital Sky Survey IV has been provided by the Alfred P. Sloan Foundation, the U.S. Department of Energy Office of Science, and the Participating Institutions. SDSS acknowledges support and resources from the Center for High-Performance Computing at the University of Utah. The SDSS web site is http://www.sdss.org. This research has made use of: the SIMBAD database, operated at Centre de Donn\'{e}es astronomiques de Strasbourg, France; the NASA/IPAC Extragalactic Database (NED) which is operated by the Jet Propulsion Laboratory, California Institute of Technology, under contract with NASA; NASA’s Astrophysics Data System; and the VizieR catalog access tool, CDS, Strasbourg, France. This research has made use of data and/or software provided by the High Energy Astrophysics Science Archive Research Center (HEASARC), which is a service of the Astrophysics Science Division at NASA/GSFC and the High Energy Astrophysics Division of the Smithsonian Astrophysical Observatory.

\bibliography{bibliography}

\end{document}